# Control of sphalerite-chalcopyrite phase transition in CdSnAs$_2$ for *n*-type thermoelectrics with high power factor

*Shoki Kishida[a], Norihiko. L. Okamoto[b], Ryoji Katsube[c], Akira Nagaoka[d], Yoshitaro Nose*[a]*

[a] Department of Materials Science and Engineering, Kyoto University, Kyoto 606-8501, Japan.

[b] Institute for Materials Research, Tohoku University, Sendai 980-8577, Japan.

[c] Department of Materials Process Engineering, Nagoya University, Nagoya 464-8603, Japan.

[d] Electrical and Electronic Engineering Program, University of Miyazaki, Miyazaki, 889-2192 Japan.

*E-mail: nose.yoshitaro.5e@kyoto-u.ac.jp

Practical applications of thermoelectric (TE) materials are constrained by less developments of high-performance *n*-type materials compared to their *p*-type counterparts. Chalcopyrite CdSnAs$_2$ is a promising *n*-type semiconductor for thermoelectrics from its narrow bandgap around 0.2 eV and exceptionally high electron mobility of $10^3$–$10^4$ cm$^2$ V$^{-1}$s$^{-1}$. In this study, we investigated the crystal growth, microstructure, and thermoelectric properties of CdSnAs$_2$. Contrary to conventional theory of unidirectional melt growth, CdSnAs$_2$ samples at higher cooling rates exhibited better crystallinity, while some cracks were observed in samples cooled more slowly. Thermal analyses clarified that a phase transition from sphalerite to chalcopyrite occurred after solidification in the case of slow cooling, leading to dislocations and cracks due to the lattice mismatch between phases. The analysis at rapid cooling suggested that supercooling lowers the solidification temperature and produces an appropriate microstructure. Consequently, the sample grown at the highest cooling rate (7.6 °C min$^{-1}$) achieved an ultrahigh power factor of 3180 μW m$^{-1}$K$^{-2}$ at 600 K and a peak *zT* of 0.62 at 682 K. In the power factor, CdSnAs$_2$ surpasses conventional binary *n*-type TE materials such as SnSe and PbTe, proving that CdSnAs$_2$ is a high potential candidate for mid-temperature TE applications.

## 1. Introduction

Thermoelectric (TE) materials, which convert temperature gradient directly into electricity, are crucial for sustainable energy generation and waste heat recovery applications.[1] TE efficiency is generally evaluated by the dimensionless figure of merit: $zT = S^2\sigma T / \kappa_{tot}$, where $S$ is Seebeck coefficient, $\sigma$ is the electrical resistivity, $T$ is absolute temperature, and $\kappa_{tot}$ is the total thermal conductivity. The total thermal conductivity is maily composed of two parts: the electronic ($\kappa_E$) and the lattice ($\kappa_L$) thermal contributions.[2] High $zT$ values indicate efficient conversion of heat into electricity, making these materials highly desirable for practical applications. Since TE materials should be combined with both n-type and p-type materials into TE devices, high $zT$ values are essential for both conduction types to maximize the overall device performance.

In addition to $zT$, another critical parameter in TE materials is the power factor, defined as $S^2\sigma$, which is directly related to the maximum power output that a material can deliver under a temperature gradient.[3] Enhancing power factor requires a balance of high electrical conductivity and a substantial Seebeck coefficient, which can be challenging as these properties are often inversely related. For practical applications, achieving a high power factor alongside an acceptable $zT$ is essential to maximize energy conversion efficiency and output, ensuring the material's competitiveness with conventional energy sources.[1]

In recent years, chalcopyrite-type TE materials have garnered significant attention due to their promising properties for energy conversion applications. High-performing materials such as $AgGaTe_2$, $CuGaTe_2$, and $CuInTe_2$ have demonstrated $zT$ values exceeding 1, highlighting their potential as efficient TE materials.[5–7] However, these high-$zT$ chalcopyrite compounds exhibit p-type conduction, leaving a gap in the availability of n-type chalcopyrite thermoelectric materials. An n-type chalcopyrite material with high power factor and $zT$ would realize TE devices based on chalcopyrite TE materials.

In that context, $CdSnAs_2$ is a promising candidate for an n-type chalcopyrite thermoelectric material, given its high electron mobility ($10^3$–$10^4$ cm$^2$ V$^{-1}$s$^{-1}$) and narrow band gap (0.2–0.3 eV). [8–10] Previous studies on the thermoelectric properties of $CdSnAs_2$ have primarily utilized hot-pressing synthesis methods, yielding peak $zT$ values of approximately 0.3 at around 700 K.[11, 12] Therefore, exploring alternative synthesis processes for $CdSnAs_2$ could unlock enhanced thermoelectric performance, making it a viable material for power generation applications.

In this study, we achieved $CdSnAs_2$ crystal growth through directional solidification using the Bridgman method. While slower cooling rates resulted in internal cracking within the crystal, faster cooling rates produced polycrystals with excellent crystallinity. Detailed thermal analyses revealed that, at slower cooling rates, a structural phase transition from sphalerite to chalcopyrite occurs after solidification, introducing strain due to lattice constant changes that impact the final microstructure. Additionally, we found that faster cooling rates mitigate this strain by supercooling, resulting in high electric conductivity. The samples grown at faster cooling rates demonstrated power factor and *zT* increases approximately three times higher than reported values synthesized by hot-pressing, achieving unprecedented TE performance among n-type chalcopyrite TE materials.

## 2. Crystal growth and characterization of CdSnAs$_2$

**Figures 1a–1c** shows the photograph of a CdSnAs$_2$ ingot after crystal growth, and wafers cut from ingots grown by the Bridgman furnace lifting with 0.75 and 75 mm h$^{-1}$, corresponding to the cooling rates of $8.8\times10^{-3}$ and 3.0 °C min$^{-1}$, respectively. The ingot length was approximately 4 cm. While cracks are observed inside the crystal grown by $8.8\times10^{-3}$ °C min$^{-1}$, no cracks formed when the cooling rate is 3.0 °C min$^{-1}$. Generally in crystal growth, slower cooling rates lead to good crystallinity, but these results show the opposite trend.

**Figures 1d** and **1e** present secondary electron images (SEI) through scanning electron microscopy (SEM) and inverse pole figure (IPF) maps obtained by electron back-scattering diffraction (EBSD) technique of these samples.[13] A large difference in the grain size is observed: the grains in the sample cooled at 3.0 °C min$^{-1}$ were more than three times smaller than those in the sample cooled at $8.8 \times 10^{-3}$ °C min$^{-1}$. This disparity arises from the greater driving force for nucleation due to supercooling at faster cooling rates, which promotes nucleation. Using energy-dispersive X-ray spectroscopy (EDS), we observed that the elemental compositions of both samples were almost the same as the stoichiometric composition (see **Table S1** in Supporting Information).

Furthermore, XRD profiles of bulk CdSnAs$_2$ samples with cooling rates of $8.8\times10^{-3}$ and 3.0 °C min$^{-1}$ are depicted in Figure 1d. The diffraction patterns of both samples can be well indexed as a chalcopyrite tetragonal CdSnAs$_2$ with a space group of $I\bar{4}2d$, and no other peak of the secondary phase appears in the patterns. These results suggest that the cooling rate affects the structure of the chalcopyrite phase after crystal growth.

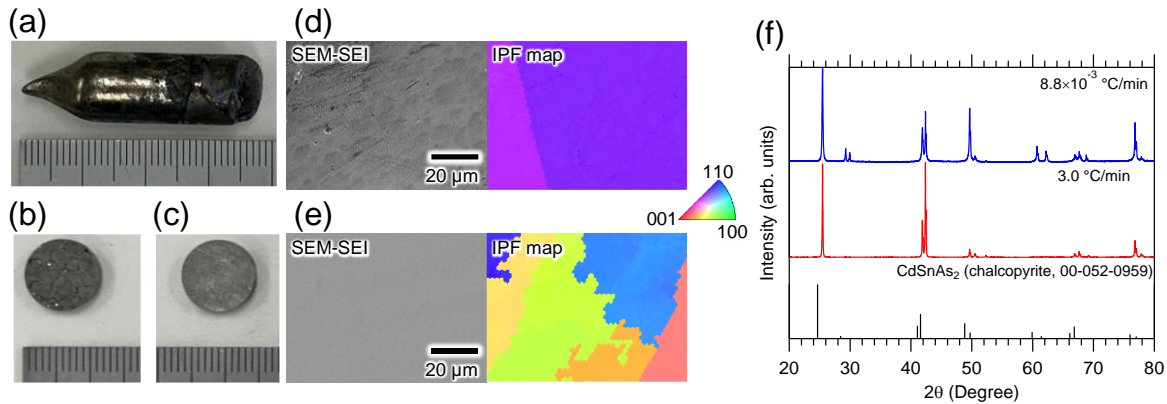

**Figure 1.** Photographs of (a) a CdSnAs$_2$ ingot, and wafers cut from ingots at the cooling rates of (b) 8.8×10$^{-3}$ and (c) 3.0 °C min$^{-1}$, with a scale bar of an mm unit. Secondary electron images (SEI) by SEM and inverse pole figure (IPF) maps for samples grown by (d) 8.8×10$^{-3}$ and (e) 3.0 °C min$^{-1}$. (f) XRD profiles of wafers shown in photos together with the reference pattern of powder specimen.

## 3. Influence of cooling rate on phase transition of CdSnAs$_2$

We conducted thermal analyses to investigate the cause of cracking during slow cooling by measuring heating and cooling curves of stoichiometric CdSnAs$_2$. The melting point of CdSnAs$_2$ was approximately 600 °C from the heating curve shown in **Figure S1** of Supporting Information, consistent with previously reported values in the literature.[12] The cooling curves and their time derivatives with the cooling rate of 2.5 and 15 °C min$^{-1}$ are illustrated in **Figures 2a** and **2b**, respectively. At the cooling rate of 2.5 °C min$^{-1}$, the derivative curve exhibits rapid increases at 596 °C and 562 °C, indicating two exothermic reactions by phase transitions. Based on the Differential Thermal Analysis (DTA) of the previous work, these reactions are attributable to solidification and structural phase transition from sphalerite ($F\bar{4}3m$, $a = 6.051$Å, **Figure 2c**) to chalcopyrite ($I\bar{4}2d$, $a = 6.099$Å, $c/a = 1.955$, **Figure 2d**), respectively. [9, 12, 14] In contrast, at a cooling rate of 15 °C min$^{-1}$, an exothermic reaction due to supercooled solidification is observed at 575 °C and the second exothermic signal due to the phase transition is not identified.

The thermal analyses indicate that crack formation is mainly caused by the structural phase transition during slow cooling. The effects of the cooling rate on the solidification and phase transition processes for the growth of CdSnAs$_2$ are discussed in **Figure 2e**. In the case of slow cooling rate, the lower driving force for solidification results in lower nucleation rates of sphalerite, allowing larger chalcopyrite grains coming from larger sphalerite grains before the phase transition from sphalerite to chalcopyrite. The mismatch of lattice parameters between the phases, $c/a$ ratio of 1.955 in chalcopyrite, may introduce defects due to lattice strain. Some cracks were actually observed as shown in **Figure 1d**. In contrast, at fast cooling rate, supercooling increases the driving force for solidification, leading to a high nucleation rate of sphalerite. Since the structural phase transition might occur before complete solidification, chalcopyrite precipitates in smaller sphalerite grains with the liquid phase coexisting, which can reduce lattice strain caused by the phase transition. In this case, the smaller chalcopyrite grains are expected due to grain size of sphalerite, which was actually observed as shown in **Figure 1e**.

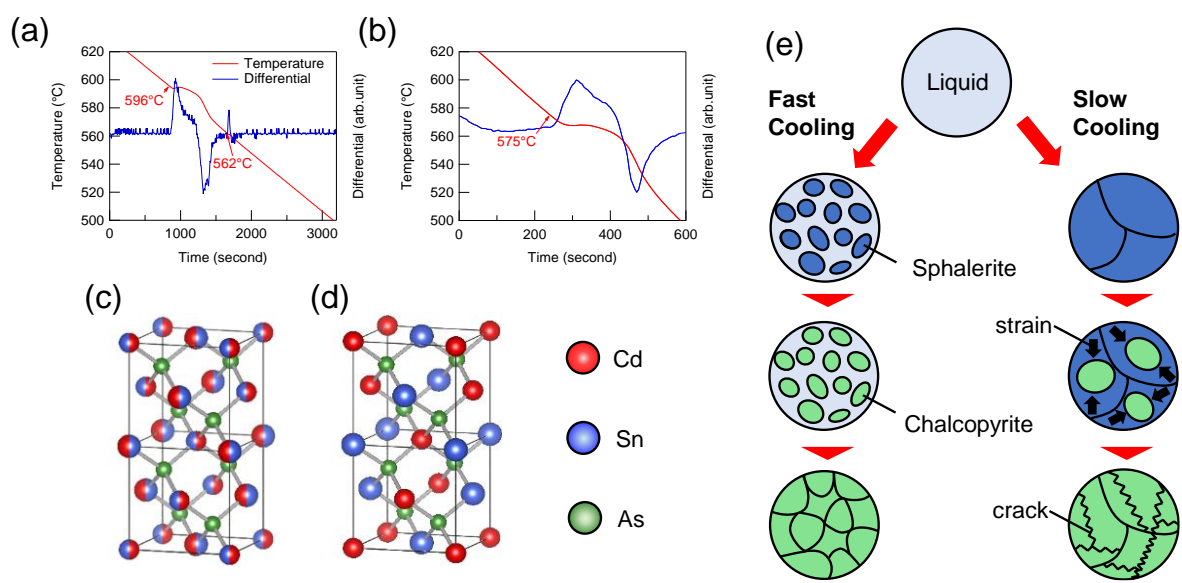

**Figure 2**. Cooling curves for CdSnAs$_2$ with the cooling rates of (a) 2.5 and (b) 15 °C/min. Red and blue lines show temperature and its time derivatives, respectively. (c) Sphalerite and (d) chalcopyrite crystal structures in CdSnAs2. In the sphalerite structure, Cd and Sn atoms occupy their sites randomly. (e) A model of microstructure formation during solidification process including sphalerite to chalcopyrite phase transition under fast and slow cooling conditions.

## 4. Microstructure, electric and thermoelectric properties with various cooling rates
### 4.1. Morphology and microstructure
To investigate the effects of cooling rate on thermoelectric properties, we conducted the crystal growth using a Bridgman furnace with lifting rates of 750, 150, 75, and 37.5 mm h$^{-1}$, corresponding to the cooling rates of 7.6, 4.4, 2.4, and 1.2 °C min$^{-1}$, respectively. SEM image of cut surface for each sample and its small area that was etched with a Br$_2$-methanol solution are presented in **Figure 3**. Observations of the cut surfaces in **Figure 3a** indicate that as the cooling rate decreases, surface damage becomes more pronounced, likely due to strain effects caused by the solid-solid phase transition. The smooth surface of etched area was only observed in the sample with a cooling rate of 7.6 °C min$^{-1}$ as shown in **Figure 3b**, where grain boundaries due to electron channeling contrast of grain orientation, and grain sizes around several tens of μm are recognized. At the cooling rates of 4.4 °C min$^{-1}$ or lower, numerous etch pits coming from dislocations are observed, indicating the introduction of strong strain from solid-solid phase transition, which might be the origin of cracks.

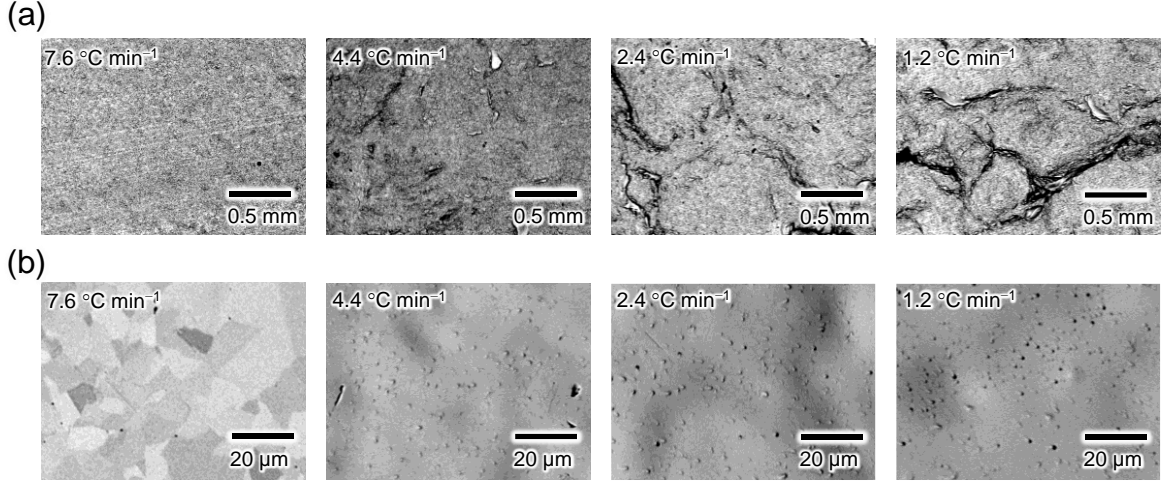

**Figure 3**. SEM images of (a) cut surface and (b) smaller areas etched with a $Br_2$-methanol solution of $CdSnAs_2$ samples grown with the cooling rates of 7.6, 4.4, 2.4 and 1.2 °C min$^{-1}$.

## 4.2. Electric and thermoelectric properties

**Figure 4a** shows the results of Hall effect measurements at room temperature for samples grown at the cooling rates of 7.6, 4.4, 2.4, and 1.2 °C min$^{-1}$. All samples exhibited a negative Hall coefficient, indicating *n*-type conduction. Moreover, the Hall mobility in these samples exceeded 4000 cm$^2$ V$^{-1}$s$^{-1}$, which is remarkably high compared to other ternary semiconductors.[15] This high mobility comes from low density of lattice defects in our crystals in addition to its low effective mass for electrons in this material ($m^* = 0.04$ $m_0$, $m_0$ is electron rest mass.) [16] and indicates high quality and homogeneity in our polycrystalline $CdSnAs_2$. Electrical conductivity increases with the cooling rate, affected by the increase in carrier concentration. To discuss the origin of increase in carrier concentration, we consider the dominant intrinsic point defects in $CdSnAs_2$. In our previous work on $ZnSnP_2$ with chalcopyrite structure, the theoretical calculation suggested that the dominant defects are antisite defects[17], and the degree of order concerning the formation of antisite defects decreases with increasing the cooling rate during crystal growth.[18] By the same analogy, it is understandable that the dominant defect is Sn antisite $Sn_{Cd}$, leading to *n*-type conduction in $CdSnAs_2$. The cooling rate dependence of electron concentration is reasonable in the viewpoint of the degree of order.

**Figures 4b** and **4c** show the temperature-dependent Seebeck coefficient and electrical conductivity, respectively. The values were continuously measured during heating and cooling cycles, and no substantial hysteresis was observed. Therefore, we adopted only the data from the heating cycle for subsequent calculations such as power factor and *zT*. The absolute value of Seebeck coefficient rises with temperature and then is saturated, which is a typical behavior in semiconductors. The optical bandgap can be estimated using the maximum value of the Seebeck coefficient and its temperature by the following equation,[19]

$$E_g \approx 2e|S_{max}|T_{max},\qquad(1)$$

where $e$ is the elementary charge. The bandgap decreases with increasing cooling rate as shown in **Figure S3** of Supporting Information. As discussed above, a higher cooling rate increases the density of antisite defects, in which donor levels form a *"band"*, leading to a reduced bandgap and higher carrier concentration. Furthermore, at higher cooling rates, the electrical conductivity shows less temperature dependence, that means a metallic electrical behavior. To discuss the differences in electrical transport due to cooling rates, we estimated the weighted mobility from the Seebeck coefficient and electrical conductivity, as shown in **Figure 5**. This metric helps discuss the average drift mobility and thermoelectric electronic transport properties.[20, 21] In general, the decrease in weighted mobility with increasing temperature is attributed to phonon scattering of charge carriers. A strong correlation was observed between weighted mobility and cooling rate. Considering the microstructure observed in **Figure 3b**, it is reasonable that lattice strain and dislocations due to the structural phase transitions contribute to electron scattering, thereby reducing electron mobility.

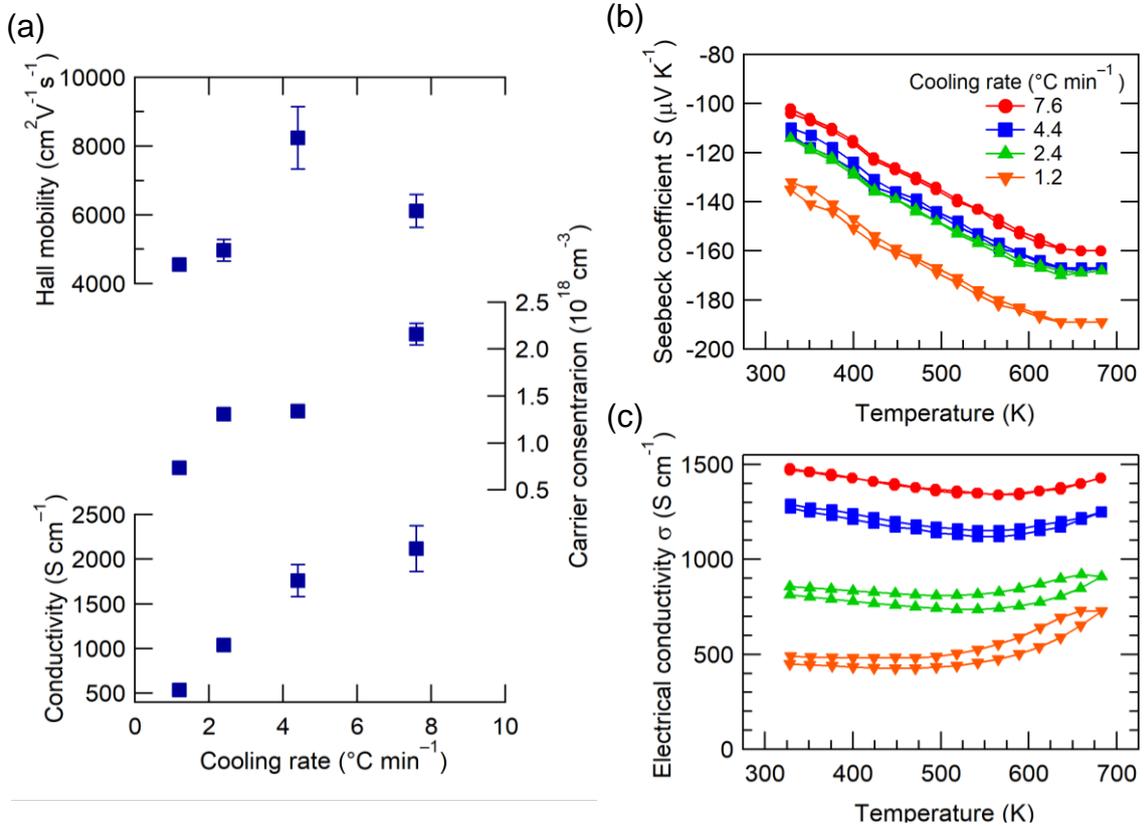

**Figure 4.** (a) Cooling rate-dependent electric properties of CdSnAs$_2$. The values represent the averages of five measurements, with the standard deviations shown as error bars. Temperature dependence of (b) Seebeck coefficient $S$ and (c) electrical conductivity $\sigma$.

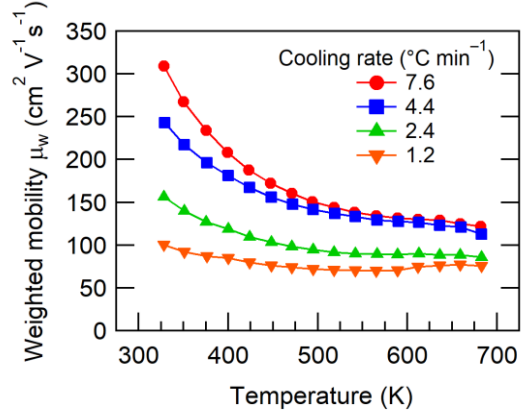

**Figure 5.** Temperature dependence of weighted mobility $\mu_w$. The weighted mobility decreases with increasing temperature due to phonon scattering.

## 4.3. Thermal transport

**Figure 6a** shows the total thermal conductivity, $\kappa_{tot}$ and lattice thermal conductivity, $\kappa_L$. The total thermal conductivity was obtained from the thermal diffusivity measured in both heating and cooling cycles. Since the hysteresis is negligible between the cycles, we adopted only heating data for the following evaluation and discussion, similar to the Seebeck coefficient and electrical conductivity. The electronic contribution to thermal conductivity, $\kappa_E$ was estimated using the Wiedemann–Franz Law, $\kappa_E = L\sigma T$. The Lorentz number $L$ (in $10^{-8}$ WΩ K$^{-2}$) was determined from the Seebeck coefficient, $S$ (in µV K$^{-1}$), following the relation[22],

$$L = 1.5 + \exp\left(-\frac{|S|}{116}\right). \qquad (2)$$

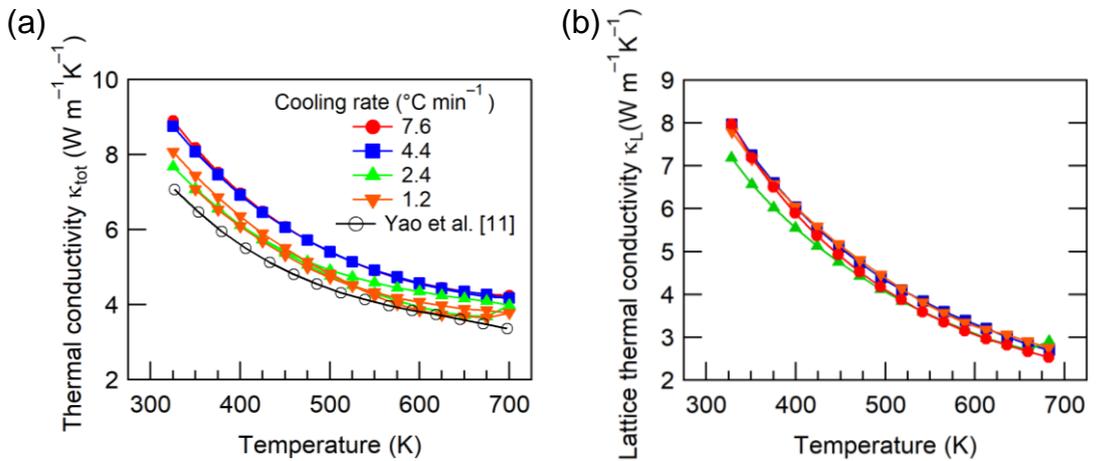

**Figure 6.** Temperature dependences of (a) total thermal conductivity $\kappa_{tot}$ and (b) lattice thermal conductivity $\kappa_L$ of CdSnAs$_2$ grown with various cooling rates. The data in the previous work[11] was obtained in samples prepared by hot-press method.

The lattice thermal conductivity was thus calculated by subtracting $\kappa_E$ from the total thermal conductivity ($\kappa_L = \kappa_{tot} - \kappa_E$) and the results are shown in **Figure 6b**. No significant difference is recognized in the thermal conductivities, in particular $\kappa_L$ is independent of cooling rates. This means that the cooling rate does not affect phonon scattering within the lattice, which may be attributed to the covalent bonding nature and strong harmonicity of CdSnAs$_2$.[12] Our crystals show higher thermal conductivity compared to samples prepared by hot-press method in the previous work[11]. Therefore, there is room to reduce the thermal conductivity by an appropriate microstructure including grain boundaries and secondary phases.

### 4.4. Quality factor, Power Factor and *zT*

The thermoelectric quality factor *B*, which determines the maximum *zT* under optimized carrier concentration, is defined as the following equation using the ratio of the weighted mobility and lattice thermal conductivity, $\mu_w/\kappa_L$ [23],

$$B = 9\frac{\mu_w}{\kappa_L}\left(\frac{T}{300}\right)^{5/2}. \tag{3}$$

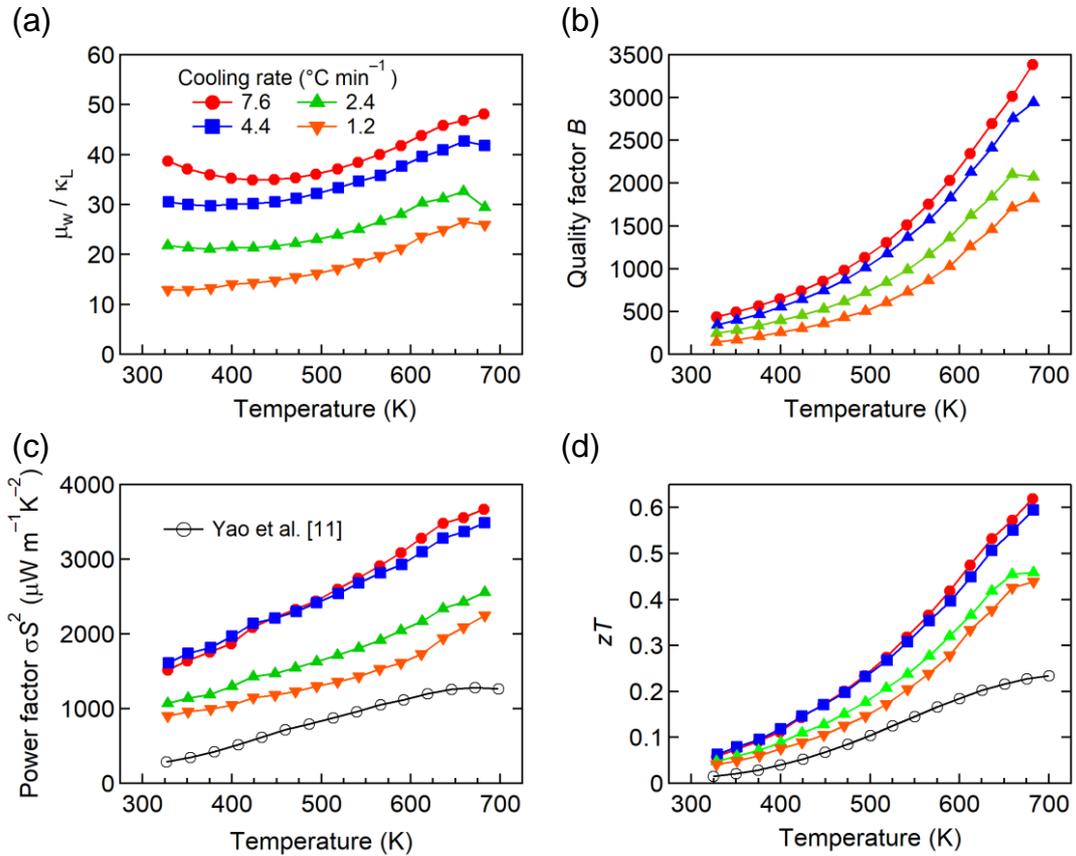

**Figure 7.** (a) Ratio of weighted mobility and lattice thermal conductivity $\mu_w/\kappa_L$, (b) quality factor *B* with the unit of cm$^2$V$^{-1}$s$^{-1}$/Wm$^{-1}$K$^{-1}$, (c) power factor $\sigma S^2$ and (d) the thermoelectric figure of merit *zT*, as a function of temperature.

The ratio $\mu_w/\kappa_L$ and the quality factor $B$ for CdSnAs$_2$ samples prepared in this work are shown in **Figures 7a** and **7b**, respectively. Both factors are higher as the cooling rate increases. In particular, the quality factor of The power factor and the figure of merit $zT$, presented in **Figures 7c** and **7d**, respectively, also increases with the cooling rate. Notably, the power factor values of 3180 μW m$^{-1}$K$^{-2}$ at 600 K at a cooling rate of 7.6 °C min$^{-1}$ are extremely high, contributing to the maximum $zT$ value of 0.62, which is 2.7 times higher than the reported value in CdSnAs$_2$. Also, this is the highest $zT$ value in any chalcopyrite compounds for *n*-type thermoelectric materials to date, including CuFeS$_2$[24] and CuInTe$_2$[25].

**Figure 8** shows power factors for various TE materials, which exhibit high values at medium temperatures, such as SnSe, PbTe and Mg$_3$Sb$_2$.[26–32] These high power factors, which were achieved without optimizing carrier concentration through doping, set a new direction for future thermoelectrics research, making CdSnAs$_2$ a promising material for intermediate-temperature power generation applications.

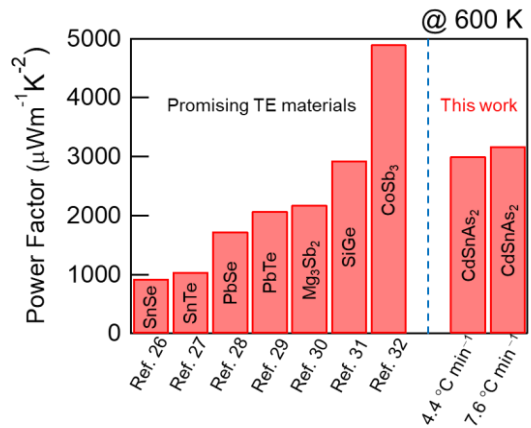

**Figure 8**. Power factors at 600 K for various TE materials[26-32] and CdSnAs$_2$ in this work: the cooling rates of 7.6 and 4.4 °C min$^{-1}$.

## 5. Conclusion

Our study reveals that the cooling rate during CdSnAs$_2$ crystal growth significantly influences the material's crystallographic structure and thermoelectric properties. While lower cooling rates lead to larger grains and lattice strain-induced cracking and dislocations due to structural phase transitions, faster cooling rates promote finer grains and reduce structural defects. Thermoelectric performance was optimized at higher cooling rates, achieving a $zT$ of 0.62 at 682 K with a cooling rate of 7.6 °C min$^{-1}$, which is the highest value in n-type chalcopyrite materials to date. This enhancement is attributed to high electrical conductivity caused by increased weighted mobility due to decreased dislocations and increased carrier concentration via Sn$_{Cd}$ antisite defects. Surprisingly, the power factor of stoichiometric CdSnAs$_2$ grown under faster cooling rates demonstrated performance comparable to leading mid-temperature thermoelectric materials like SnSe and PbTe, positioning it as a promising candidate for thermoelectric applications in the intermediate temperature range.

## 6. Experimental Section

*Crystal growth*: In total 15 g of raw elements, including Cd grains (99.9999%, Kojundo Chemical Laboratory), Sn grains (99.99%, My Assoc) and As grains (99.9999%, Kojundo Chemical Laboratory) were weighted according to stoichiometry and placed in carbon-coated quartz ampules with an 11 mm inner diameter × 2 mm thickness. Sn and As grains were chemically etched using 0.1 M HCl solution and 1 M $HNO_3$, respectively, prior to sealing in the ampules. The ampules were evacuated under the pressure of $10^{-2}$ Pa, and approximately $3.0 \times 10^4$ Pa of $N_2$ gas was introduced into the ampules before the ampules were sealed. Then, the quartz ampules were set in a Bridgman furnace with three heaters to precisely control the temperature gradient. The sealed ampules were located at the position around 610 °C for homogenization. The furnace was raised, and the samples were unidirectionally solidified from the bottom. The crystal growth of $CdSnAs_2$ was carried out with varied cooling rates ($8.8 \times 10^{-3}$, 1.2, 2.4, 3.0, 4.4, 7.6 °C $min^{-1}$) by changing the lifting speed of the furnace. The temperature of the bottom of the ampule was monitored with a K-type thermocouple during crystal growth, and the temperature-time relationships during cooling from 605 °C to 595 °C were used to estimate the cooling rate of each sample.

*Characterization*: The microstructures of the $CdSnAs_2$ crystals were analyzed by electron backscatter diffraction (EBSD, DigiView, EDAX), which was integrated into a field emission scanning electron microscopy (FE-SEM, JSM-7001FA, JEOL) system. The samples for the EBSD analysis were milled using an Ar ion beam processing system (CROSS SECTION POLISHER"TM", IB-19520CCP, JEOL) with an accelerating voltage of 5 kV. The obtained phase of the grown crystals analyzed by X-ray diffraction (XRD, X'Pert Pro, Panalytical) and the composition of the crystals was analyzed by a scanning electron microscope (SEM, JCM-6000PLUS, JEOL) equipped with an energy dispersive X-ray spectrometer (EDS, JED-2300, JEOL). In addition, the sample morphology was analyzed by performing SEM observations on the surface of $CdSnAs_2$ after chemical etching. The etching was conducted using a $Br_2$/methanol solution at 40 °C with stirring at 150 rpm for 5 minutes.

*Thermal analysis*: In total 1 g of raw elements, including Cd, Sn and As grains were weighted according to stoichiometry and placed into a quartz ampule with a 7 mm inner diameter × 2 mm thickness. The ampule was evacuated under the pressure of $10^{-2}$ Pa, and approximately $3.0 \times 10^4$

Pa of $N_2$ gas was introduced into the ampule before the ampules were sealed. Then, a K-type thermocouple was attached to the sealed ampule using stainless steel wires to monitor the temperature of the bottom of the ampule during heat treatment. The temperature curves were obtained by heating and cooling the samples in a box furnace at various constant cooling rates. We confirmed the thermal reaction by taking the time derivative of the obtained temperature curve.

*Thermoelectric Measurement*: The Hall measurement was accomplished by Van der Pauw method in a Hall effect measurement system (Resitest 8300, Toyo Technica) under a 0.38 T magnetic field at room temperature. The electrical conductivity and the Seebeck coefficient were measured simultaneously in a helium atmosphere using a commercial system (ZEM2, ADVANCE RIKO) on samples with dimensions of approximately $2 \times 2 \times 6$ mm$^3$ at 325–700 K. The thermal conductivity κ was calculated based on the equation: $\kappa = DC_p\lambda$. The density $D$ was determined by Archimedes' method, and the specific heat $C_p$ was derived using the Dulong-Petit model, $C_p = 3n\text{R}$, where $n$ is the number of atoms per formula unit and R is the gas constant. The thermal diffusivity λ was measured using a laser flash analyzer (LFA467, Netzsch) on samples with dimensions of approximately 2 mm × 6 mm × 6 mm in an argon atmosphere at 325–700 K. This measurement was carried out after the samples were coated with a thin graphite layer to minimize the errors caused by the material emissivity.

**Supporting Information**

Supporting Information is available on the web and from the author.

**Acknowledgements**

This work was financially supported by JSPS KAKENHI (Nos. 23H01739 and 23K26432) and Japan Organization for Metals and Energy Security (JOGMEC). The authors thank Prof. G. J. Snyder (Northwestern University) for fruitful discussions.

# Supporting Information for:
# Control of sphalerite-chalcopyrite phase transition in CdSnAs₂ for *n*-type thermoelectrics with high power factor

*Shoki Kishida[a], Norihiko. L. Okamoto[b], Ryoji Katsube[c], Akira Nagaoka[d], Yoshitaro Nose*[a]*

[a] Department of Materials Science and Engineering, Kyoto University, Kyoto 606-8501, Japan.

[b] Institute for Materials Research, Tohoku University, Sendai 980-8577, Japan.

[c] Department of Materials Process Engineering, Nagoya University, Nagoya 464-8603, Japan.

[d] Electrical and Electronic Engineering Program, University of Miyazaki, Miyazaki, 889-2192 Japan.

*E-mail: nose.yoshitaro.5e@kyoto-u.ac.jp

**Table S1**. Composition of grown CdSnAs₂ by SEM-EDS

| Cooling rate (°C min⁻¹) | Composition (mol%) | | |
|---|---|---|---|
| | As | Cd | Sn |
| 3.0 | 50.7 | 24.8 | 24.4 |
| 8.8×10⁻³ | 50.8 | 24.8 | 24.4 |

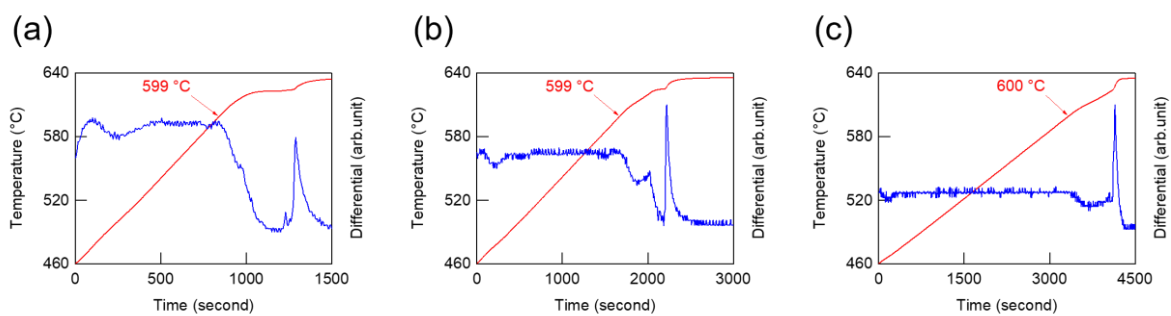

**Figure S1**. Heating curves for CdSnAs₂ with the heating rates of (a) 10, (b) 5 and (c) 2.5 °C min⁻¹. Red and blue lines show the temperature and its time derivative, respectively. These curves consistently show the exothermic reactions due to melting at around 600 °C.

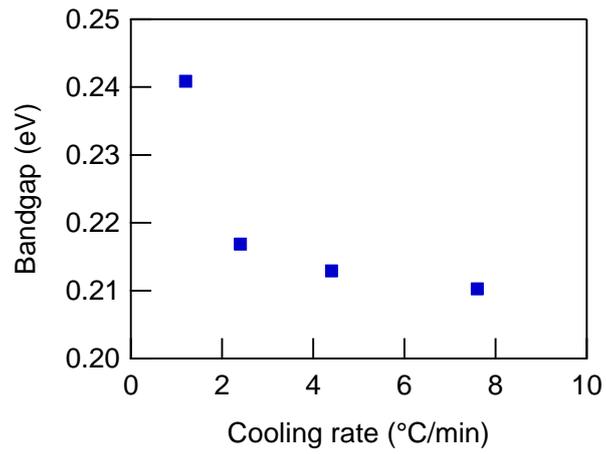

**Figure S2**. Relationship between bandgap and cooling rate in CdSnAs$_2$ crystals, calculated by the Goldsmid−Sharp relation.